\documentclass[11pt]{article}

\usepackage[letterpaper,top=2cm,bottom=2cm,left=2cm,right=2cm,marginparwidth=1.75cm]{geometry}

\usepackage{graphicx}
\usepackage{subcaption}
\usepackage{booktabs}
\usepackage{amsmath}
\usepackage{bm}
\usepackage{tocloft}
\usepackage{blindtext}
\usepackage{natbib}
\usepackage[colorlinks=true, allcolors=blue]{hyperref}

\title{Comparing estimators of discriminative performance of time-to-event models}
\author{Ying Jin, Andrew Leroux}

\begin{document}


\maketitle

\begin{abstract}
Predicting the timing and occurrence of events is a major focus of data science applications, especially in the context of biomedical research. Performance for models estimating these outcomes, often referred to as time-to-event or survival outcomes, is frequently summarized using measures of discrimination, in particular time-dependent AUC and concordance. Many estimators for these quantities have been proposed which can be broadly categorized as either semi-parametric estimators or non-parametric estimators. In this paper, we review various estimators' mathematical construction and compare the behavior of the two classes of estimators. Importantly, we identify a previously unknown feature of the class of semi-parametric estimators that can result in vastly over-optimistic out-of-sample estimation of discriminative performance in common applied tasks. Although these semi-parametric estimators are popular in practice, the phenomenon we identify here suggests this class of estimators may be inappropriate for use in model assessment and selection based on out-of-sample evaluation criteria. This is due to the semi-parametric estimators' bias \textit{in favor} of models that are overfit when using out-of-sample prediction criteria (e.g., cross validation). Non-parametric estimators, which do not exhibit this behavior, are highly variable for local discrimination. We propose to address the high variability problem through penalized regression splines smoothing. The behavior of various estimators of time-dependent AUC and concordance are illustrated via a simulation study using two different mechanisms that produce over-optimistic out-of-sample estimates using semi-parametric estimators. Estimators are further compared using a case study using data from the National Health and Nutrition Examination Survey (NHANES) 2011-2014. 
\end{abstract}



\section{Introduction}
\label{s:intro}

Modeling time-to-event outcomes, also known as survival analysis, is a major area of methodological development in statistics and machine learning and is relevant to many applied data science tasks. Broadly, the prediction accuracy of time-to-event outcomes can be assessed locally (i.e., for a fixed time point) or globally (summarized over a set of time points). Local performance is often assessed by time-dependent Receiver Operating Characteristic Curve (ROC) and area under the ROC curve (AUC), while global performance is often assessed by concordance (C-index). In this paper, we review the formulation of these estimators, compare their behavior in realistic data science scenarios, and identify the cause of undesirable out-of-sample behavior in semi-parametric estimators of discrimination. We restrict our focus to the Cox proportional hazards model framework \citep{cox1972}, one of the fundamental statistical approaches to modeling time-to-event outcomes \citep{Harrell1996, ElHafeez2021}. The accuracy and behavior of estimators of discrimination has been studied in previous literature, though the focus has been on the effect of violations of the proportional hazards assumption. Specifically, \cite{VANGELOVEN2021107095} found that the semi-parametric estimator of time-dependent AUC can be biased when the proportional hazard assumption does not hold. Other work by \cite{Blanche2018} found that Harrell's c-index can overestimate discriminative performance of biomarkers for t-year risk in such scenarios. While violations of the proportional hazards assumption are an important consideration, they are not the only source of potential issues relating to the validity of estimators of discrimination. The behavior of estimators which we discuss in this paper is wholly different from the effect of such model misspecification and applies to correctly specified models in which the proportional hazards assumption holds.

Some semi-parametric estimators \citep{hz2005, SongZhou2008} are consistent under the assumption of proportional hazard and independent censoring. However, a previously unidentified flaw exists for a specific class of estimators, which renders them inappropriate for use in many data science contexts where out-of-sample accuracy is used to perform model selection \citep{Yates2022, Burman1989}. Specifically, the semi-parametric estimators considered in the current work have the potential to substantially overestimate out-of-sample discriminative performance, even when the model is correctly specified. This poor behavior is most easily seen in the context of 1) model overfit and 2) covariate misalignment. The latter we define to mean situations where the covariates of the \textit{test} sample have a different distribution from the data used to \textit{train} the model. This issue of covariate misalignment can cause poor out-of-sample behavior of semi-parametric estimators of discrimination even when the model-predicted risk in the test sample is well calibrated. 

\begin{figure}
\centering
\includegraphics[width=\textwidth]{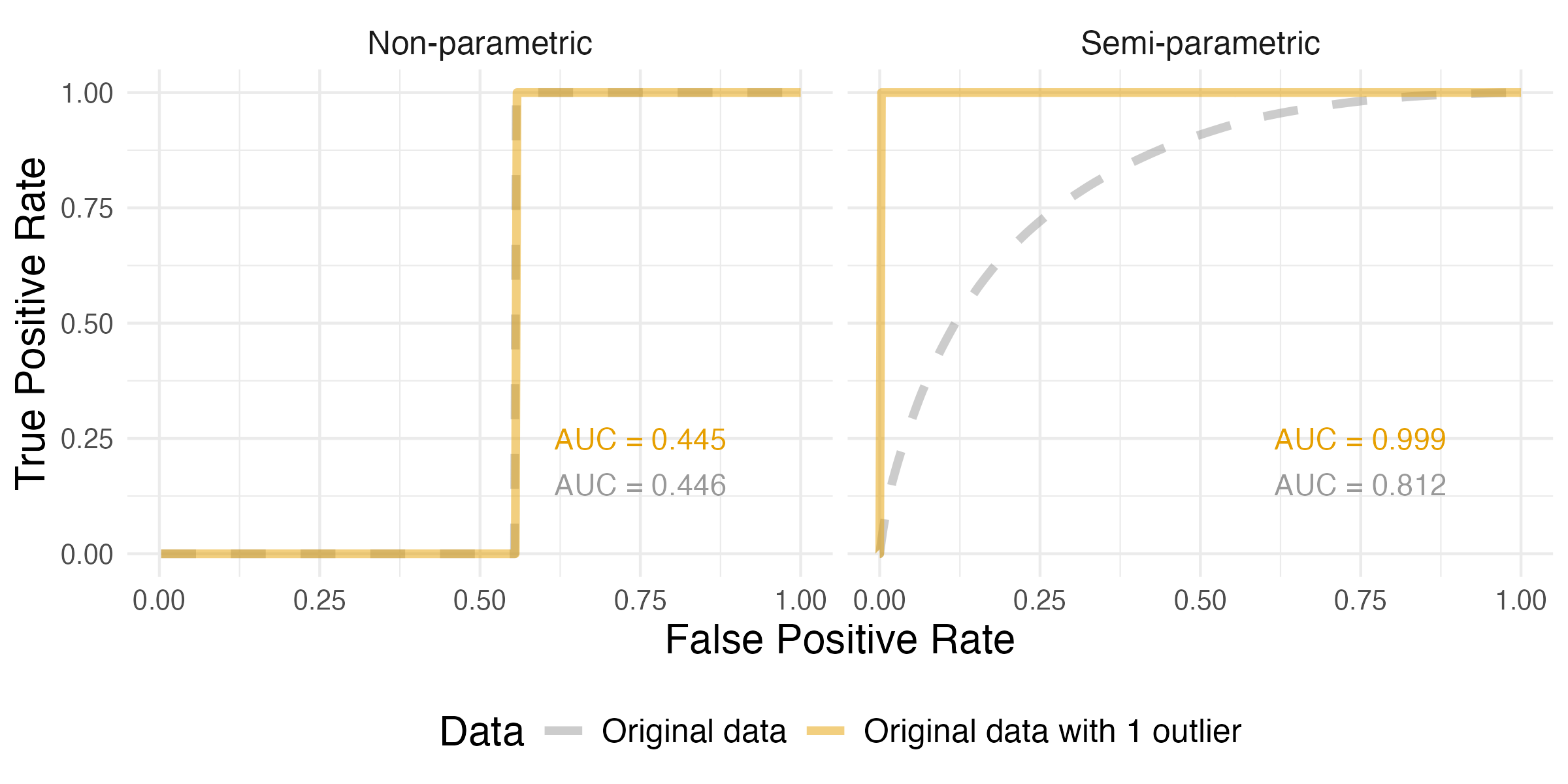}
\caption{Change of out-of-sample semi-parametric estimation of discriminative performance after introducing one outlier to the testing sample (N=500). The datasets represented by the yellow and grey lines are identical, except for one outlier with outlying values of covariates.}
\label{fig-outlier}
\end{figure}

The use of out-of-sample prediction accuracy as the gold standard for model selection and assessment is done in large part to avoid the tendency of in-sample estimates to be overly optimistic due to model overfit to the training data \citep{Yates2022, Arlot2010}, allowing for greater generalizability of predictions. As a result, it is important for out-of-sample estimators to accurately identify when a model is either overfit to the data or provides poorly calibrated risk predictions. To understand the potential magnitude of the phenomena we've identified, we provide an illustration using a case of covariate misalignment via the presence of an outlier in the training data. Figure \ref{fig-outlier} provides an illustration of out-of-sample overestimation under covariate misalignment caused by the presence of one outlier, where we evaluate the out-of-sample performance of the same Cox regression model on two different datasets using the Incident/Dynamic AUC \citep{hz2005} at a particular follow-up time ($t=0.27$). Specifically, Figure~\ref{fig-outlier} presents estimated Incident/Dynamic ROC curves, the integral of which is Incident/Dynamic AUC. Values of Incident/Dynamic AUC close to 1 indicate near-perfect discrimination (or, equivalently, ROC curves which tend toward the upper left quadrant of the plot), while a value close to 0.5 (ROC curves near the identity line) indicates the predictor is no more prognostic than a coin flip. The Cox model was correctly specified and fit to a training dataset with 300 subjects, with 228 subjects at risk at $t = 0.27$. Out-of-sample performance is evaluated on two testing datasets that are identical except for one subject: the dataset represented by the yellow line introduced one outlier with abnormally large values of covariates. Both testing sets have a sample size of 500. At the t = 0.27, the number of subjects at risk is 364 in the dataset without the outlier and 365 in the dataset with the outlier. As Figure \ref{fig-outlier} shows, this one single observation has driven out-of-sample semi-parametric AUC from 0.812 to 0.999 at this time point. Intuitively, a single observation out of 365, whether their risk was \textit{accurately} predicted by the model or not, should not shift AUC by such a large margin. Moreover, this individual's risk was poorly predicted by the model, suggesting discrimination should \textit{decrease} rather than substantially increase to near \textit{perfect} discrimination. That is, the estimated log hazard of this subject is 3.6 times higher than the second largest log hazard in the same sample at this time point, with a predicted probability of survival beyond time $t=0.27$ of essentially $0$. However, the subject experienced the event at time 0.85, indicating the Cox model has poorly estimated the observation's true risk. In this example, the semi-parametric estimator of Incident/Dynamic AUC failed to properly reflect model performance or generalizability. The poor behavior of this estimator of time-dependent AUC at some time $t$, $\text{AUC}(t)$, is caused by the disconnection between the estimator (which uses the magnitude of model estimated risk in it's construction) and the actual event status of samples in the risk set who experience an event beyond the current time (i.e., for some $t^* > t$). We provide additional details in Section~\ref{subsec:mach_sim}. 

Non-parametric estimators, on the other hand, do not exhibit this behavior as the estimators do not include the \textit{magnitude} of the estimated risk and instead consider only the \textit{relative ranking} of risk. However, non-parametric estimators of time-dependent AUC exhibit higher variability, frequently with large jumps between neighboring time points. This is because at each event time point, the number of events is usually low. Only one or a few events would be observed, causing the sensitivity estimates to be highly unstable, even fluctuate between extreme values as 0 and 1. As shown in the left panel of Figure \ref{fig-outlier}, the non-parametric incident/dynamic ROC curve at this time point is a step function, jumping from 0 to 1 when the cutoff value of risk score reaches a certain point. The resulting AUC at this point is low (0.44), but stays unchanged after introducing an extreme outlier. Therefore, these estimators often requires additional smoothing. For example, \citet{Shen2015} explored fractional polynomial regression for smoothing. Both \citet{Song2012} and \citet{Saha-Chaudhuri2012} use kernel functions for smoothing. In this paper, we address this variability issue by smoothing the estimates over time using penalized regression splines \citep{wood2004}

Other work looking at properties of  estimators of discrimination has focused on the assumptions of proportional hazard and independent censoring. As examples, \cite{VANGELOVEN2021107095} showed the Incident/Dynamic AUC estimator may be biased when the proportional hazard assumption is violated. \cite{Schmid2012} compared different concordance estimators through an extensive simulation study under the violation of either assumptions, and found that the estimator proposed by \cite{hz2005} showed bias in both cases. However, the behavior of semi-parametric estimators we describe in our work here is seen even when the conditions above are satisfied, and represents a different and heretofore unidentified feature. Although compelling arguments have been made that discrimination is an inappropriate criteria for model selection in the context of time-to-event analyses, such measures provide one piece of useful information about the predictions made by a particular model, thus have been and continue to be used frequently for model assessment and selection. For example, \cite{Cornec-LeGall2016} used the semi-parametric estimator of AUC proposed by \cite{hz2005} to predict renal survival in Autosomal Dominant Polycystic Kidney Disease. The non-parametric Harrell's C index was used in \cite{Stephenseon2005} to predict the 10-year risk of recurrence of prostate cancer after radical prostatectomy, and in \cite{Pencica2009} for 30-year risk of cardiovascular disease. Due to their utilization by practitioners, understanding the properties and shortcomings of various estimators is critical. Thus, we add to the literature by identifying the pros and cons of different estimators. We hope to provide appropriate references for investigators who need these measures for tasks such as model selection, model assessment, etc. 

In the following section, we define the evaluation metrics of discriminative performance for time-to-event models, such as Incident/Dynamic AUC and concordance. Section \ref{s:estimator} then introduces their estimators and identifies the source of undesirable behaviors. A simulation study in Section~\ref{sec:simulation} compares the behavior of different classes of estimators in the context of model overfit and covariate misalignment. Section~\ref{s:nhanes_data} further illustrates their practical utility with an application to the 2011-2014 National Health and Nutrition Examination Survey (NHANES, 2011-2014) data. 

\section{Measures of discriminative Performance}
\label{s:measure}


For risk prediction in time-to-event models, an estimate of risk (or, more precisely, a ranking of risk) is required for all units/subjects at a set of times for the range of the time-to-event random variable. Given these estimates of risk, local discrimination involves defining a set of "cases" and "controls" specific to time $t$ and then calculating the corresponding ROC curve at this time point using the time-specific risk estimates. This "local" AUC is broadly referred to as time-dependent AUC, with two popular estimands being Incident/Dynamic AUC and Cumulative/Dynamic AUC \citep{hz2005}. The former compares discrimination of risk for incident cases (events observed at a specific time point) to dynamic controls (subjects at risk at the same time point). It is often used when researchers are interested in makers' discriminative ability with temporal updates, such as dynamic prediction/forcasting \citep{Cornec-LeGall2016, VANGELOVEN2021107095}. The use of incident cases prevents redundant information over time \citep{Blanche2013_2}. The latter focuses on historical cases (events observed up to a specific time point) and dynamic controls, which is often used for prognosis from baseline \citep{Mortensen2017}. \cite{Blanche2013_2} also suggested its use for clinical decision-making, such as enrollment in clinical trials. Global summaries of discrimination look at how well the risk predictions discriminate for across a range of (or all) time points, $t$, with Concordance or C-index \citep{Harrell1996, Uno2011} being common measures. Both local and global measures assess how well risk rankings compare to observed event times. The focus of this paper is restricted to the local measure of Incident/Dynamic AUC ($\text{AUC}^{I/D}(t)$) and the global measure of Concordance ($C$), a weighted average of $\text{AUC}^{I/D}(t)$ over time. 

Let $i = 1,\ldots, N$ denote an individual for whom we observe a time-to-event outcome,  $T_i^*$,  subject to right censoring, denoted as $C_i$. For each individual, we observe $[T_i, \delta_i, \mathbf{X}_i^t]$ where $T_i = min(T_i^*, C_i)$ is the observed time (minimum of censoring time $C_i$ and true event time $T_i^*$). $\delta_i = 1(T_i^* \leq C_i)$ is the event indicator, and  $\mathbf{X}_i \in R^p$ is a vector of time-fixed covariates. Censoring time $C_i$ is assumed to be independent of event time $T_i^*$ conditional on $\mathbf{X}_i$. Additionally, the data is assumed to be generated from a proportional hazards model \citep{cox1972}, where the log hazard of the event time takes on the additive form
\begin{equation}
\log \lambda(t|\mathbf{X}_i) = \log \lambda_0(t) + \mathbf{X}_i^t \boldsymbol{\beta}\
= \log\lambda_0(t) + \eta_i; \hspace{0.5cm} t > 0
\end{equation}

In the proportional hazards model, $\log \lambda(t|\mathbf{X}_i)$ is the conditional log-hazard for subject $i$ given their covariate vector $\mathbf{X}_i$, and $\log \lambda_0(t)$ is the log baseline hazard which is shared across the population and left unspecified. The vector of regression coefficients, $\boldsymbol{\beta}$, is unknown and corresponds to the linear contribution of each element of $\mathbf{X}_i$ to the log hazard. Finally, $\eta_i$ is the overall contribution of covariates to log hazard, indicating the subject-specific deviation of log hazard from the population level. Hereafter we refer to $\eta_i$ as the risk score of subject $i$. Note that the results and findings below readily extend to extensions of the classical Cox model where risk depends on time ($\eta_i(t)$) through either a time-specific coefficient ($\beta(t)$), a time-specific covariate ($\mathbf{X}_i(t)$), or both.  

\subsection{Incident/Dynamic AUC}
\label{subsec:define_auc}

Incident/Dynamic AUC, or $\text{AUC}^{I/D}(t)$ \citep{hz2005}, generalizes the notion of AUC for binary data, allowing for time-dependent discrimination for time-to-event models. At a specific time t, $\text{AUC}^{I/D}(t)$ is achieved by calculating the incident sensitivity ($\text{sensitivity}^{I}(c,t)$) and dynamic specificity ($\text{specificity}^{D}(c,t)$) at a series of unique thresholds $c$ for risk score $\eta_i$, deriving a time-specific ROC curve and estimating the area under it. As with AUC for binary data, $\text{AUC}^{I/D}(t) \in [0,1]$, with values closer to $1$ indicating better discrimination and values near $0.5$ indicating the risk score $\eta_i$ is no better at discriminating events at time $t$ than a flip of a coin. We describe these estimands in more detail below. 

Incident sensitivity and dynamic specificity are defined as
\begin{equation}
    \text{sensitivity}^{I}(c,t) = {TP}_t^{I}(c) = Pr(\eta_i > c|T_i^* = t)
\end{equation}
\begin{equation}
    \text{specificity}^{D}(c,t) = 1-FP_t^{D}(c) = Pr(\eta_i \leq c|T_i^* > t) 
\end{equation}

where ${TP}_t^{I}(c)$ and $FP_t^{D}(c)$ are abbreviations for time-specific incident true-positive and dynamic false-positive rate, respectively, and $c$ is a threshold of risk score $\eta_i$. Using the above definitions for incident sensitivity and dynamic specificity, we can then define the Incident/Dynamic ROC curve. Let p denote the value of $FP_t^{D}(c)$, then 
\[ \text{ROC}_t^{I/D}(p) = TP_t^{I}\{[FP_t^{D}]^{-1}(p)\} \]

From which it follows that $\text{AUC}^{I/D}(t)$: 
\[ \text{AUC}^{I/D}(t) = \int_0^1 \text{ROC}_t^{I/D}(p)dp \]

In practice, $\text{AUC}_t^{I/D}(t)$ is generally approximated by numeric integration, evaluating \\$[TP_t^{I}(c), FP_t^{D}(c)]$ at every unique value of risk score among the risk set at a specific event time. Hereafter, we will denote the risk set at a specific time t as $R(t) = \{i: T_i \geq t\}$. The collection of unique values of risk score $\eta$ among subjects in $R(t)$ will be denoted as $\boldsymbol{\eta}_{R(t)} = \cup_{i\in R(t)} \{\eta_i\}$. 

\subsection{Concordance}
\label{subsec:define_c}

Concordance, defined as $C = Pr(\eta_i < \eta_j| T_i^* > T_j^*)$ for a randomly selected pair $(i,j)$, represents the overall agreement between true event times and risk scores. As with $\text{AUC}^{I/D}(t)$, $C \in [0,1]$, with values closer to $1$ denoting better global discrimination of the risk score. In practice, $T_i^*$ may have support beyond the duration of a study, resulting in a need to administratively censor participants at some follow-up time $\tau$ (e.g., the end of the study). In the context of administrative censoring, the estimand becomes $C^\tau = Pr(\eta_i < \eta_j| T_i^* > T_j^*, T_j^* < \tau)$. It has been shown that this truncated concordance is a weighted average of Incident/Dynamic AUC \citep{hz2005}:
\begin{equation}
C^\tau = \int_0^{\tau} 
\text{AUC}^{I/D}(t)w^{\tau}(t)dt;
\hspace{0.5cm}
w^{\tau}(t) = \frac{2f(t)S(t)}{1-S^2(\tau)} \label{eq:concordance}
\end{equation}
where $S(t)$ is the marginal survival function (not conditional on covariates) and $f(t)$ is the marginal probability density function of time to event.

\section{Estimators of discriminative Performance}
\label{s:estimator}

In this section, we discuss methods of estimating $\text{AUC}^{I/D}(t)$ and $C^\tau$ defined in Section~\ref{s:measure} above. We distinguish between semi- and non-parametric estimators as the different classes of estimators with regard to both their formulation and out-of-sample behavior.

\subsection{Incident/Dynamic AUC} 
\label{subsec:est_auc}

Recall, $\text{AUC}^{I/D}(t) = \int_0^1 \text{ROC}_t^{I/D}(p)dp$. This integral can be approximated numerically with $\sum_p \delta_p \hat{\text{ROC}}^{I/D}(p)$, where p is the estimated dynamic false positive rate at every unique value of risk score at this time: $p=\hat{\text{FP}}^D(c)$, $ c \in 
\boldsymbol{\eta}_R(t) \cup -\infty$. $\delta_p$ here is a quadrature weight typically determined using the trapezoid rule. Evaluating at $c = -\infty$ ensures the estimated $\text{ROC}_t^{I/D}$ passes through the point $(1,1)$. Thus, different estimators of $\text{AUC}^{I/D}(t)$ arise from the use of different estimators of dynamic specificity and/or incident sensitivity. Here, all estimators considered use the same non-parametric estimator of dynamic specificity but differ in their approach to estimating incident sensitivity. 

Specifically, suppose we have obtained estimated coefficient $\hat{\boldsymbol{\beta}}$ and used it to estimate individual risk scores $\hat{\eta}_i = \boldsymbol{X}_i^t\hat{\boldsymbol{\beta}}$.  Dynamic specificity can then be estimated 
 as follows: 
\begin{equation}
1-\widehat{\text{specificity}}^{D}(c,t)= \hat{FP}_t(c) = 
\frac{\sum_{k}I(\hat{\eta}_k>c)I(T_k>t)}{\sum_{j}I(T_j>t)}
\label{eq:spec_np} 
\end{equation}

This estimator of dynamic false-positive rate is built based on the plug-in principle, counting up the proportion of individuals with an estimated risk greater than a particular threshold among those who have not experienced the event by time $t$.

Moving on to estimators of incident sensitivity, first consider a non-parametric estimator based similarly on the plug-in principle:

\begin{equation}
\widehat{\text{sensitivity}}^{I}(c,t)=
\hat{TP_t}^{NP}(c)=  \frac{\sum_{k}I(\hat{\eta}_k>c)I(T_k=t)I(\delta_k=1)}{\sum_{j}I(T_j=t)I(\delta_j=1)}    
\label{eq:sens_np} 
\end{equation}

This non-parametric estimator is inherently more variable than the non-parametric estimator of dynamic specificity in Eq.\ref{eq:spec_np}. To see this, note that Eq.\ref{eq:sens_np} is based on counting the proportion of individuals who have a risk score above a particular threshold $c$ among those with an observed event at $t$. As such, the resulting time-specific ROC curve would be a step function with a single step (See Figure~\ref{fig-outlier} left panel). The estimator of  $\text{AUC}^{I/D}(t)$ obtained by non-parametric specificity (Eq.\ref{eq:spec_np}) and sensitivity (Eq.\ref{eq:sens_np}) is therefore a non-parametric estimator.

Next, consider the semi-parametric estimator of $\text{AUC}^{I/D}(t)$ proposed by \cite{hz2005}. It uses the same non-parametric estimator of $\hat{FP}_t(c)$ in Eq.\ref{eq:spec_np}, but differs in their estimator of incident sensitivity:

\begin{equation}
\hat{TP_t}^{SP}(c)=\frac{\sum_{k}I(\hat{\eta}_k>c)I(T_k\geq t)exp(\hat{\eta}_k)}{\sum_{j}I(T_j\geq t)exp(\hat{\eta}_j)}
\label{eq:sens_sp}
\end{equation}

Instead of counting the proportion of true-positive subjects, this estimator conditions on all subjects at risk at time $t$ and weigh the subjects by their exponential estimates of risk score. The subject-specific weight $\frac{exp(\hat{\eta}_k)}{\sum_{j}I(T_j\geq t)exp(\hat{\eta}_j)} $ depends on the \textbf{values} coefficient estimates $\hat{\boldsymbol{\beta}}$, thus \textbf{parametric} in nature. In addition, we note the lack of dependence on actual observed events at $t$ (i.e. $\delta_i$ appears nowhere in this formula). These two points are the key marks to distinguish the formulation of semi-parametric estimators from the non-parametric ones. They are also the reasons that semi-parametric estimators, though consistent under our proposed framework \citep{10.2307/2680613} and relatively smooth in practice, suffer from over-optimistic inflation of out-of-sample discrimination. We return to this point in Section~\ref{subsec:mach_sim}.

\subsection{Concordance}
\label{par:est_c}

Truncated concordance, $C^{\tau}$ (Eq.\ref{eq:concordance}) may be estimated using the result linking $C^{\tau}$ to $\text{AUC}^{I/D}(t)$: $\text{AUC}^{I/D}(t) = \int_0^1 \text{ROC}_t^{I/D}(p)dp$, or using other semi- and non-parametric estimators. First consider estimating $C^{\tau}$ as the weighted integral of $\text{AUC}^{I/D}(t)$. This can be done using any estimator of  $\text{AUC}^{I/D}(t)$, with weights derived from estimated marginal survival function: $\hat{w^{\tau}}(t) = \frac{2\hat{f}(t)\hat{S}(t)}{1-\hat{S}^2(\tau)}$. 

However, as was mentioned previously, the non-parametric estimator of $\text{AUC}^{I/D}(t)$ derived from non-parametric specificity (Eq.\ref{eq:spec_np}) and sensitivity (Eq.\ref{eq:sens_np}) is highly variable, which presents a challenge for numeric integration.
Some previous literature has approached this issue by smoothing the $\text{AUC}^{I/D}(t)$ estimates, using lowess, kernel \citep{Song2012} or fractional polynomial\citep{Shen2015}. We therefore propose a complementary approach. Specifically, we propose to smooth the non-parametric $\hat{\text{AUC}}^{I/D}(t)$ using penalized regression splines via the \textit{mgcv} package \citep{wood2003, wood2011, wood2017} in \textit{R} \citep{Rsoftware}. That is, we estimate the additive model
\[
    \hat{\text{AUC}}^{I/D}(t) = 
    \tilde{\text{AUC}}^{I/D}(t) + \epsilon(t) =
    \sum_{k=1}^K \xi_k B_k(t) + \epsilon(t)
\]

Here $\tilde{\text{AUC}}^{I/D}(t)$ is the smoothed non-parametric Incident/Dynamic AUC estimates, modelled as the linear combination of a set of spline basis functions $B_1(t)...B_K(t)$ subject to penalty on second derivative. $\epsilon(t)$ denotes a random Gaussian noise that is independent and identically distributed.

Please note that the weight estimator $\hat{w^{\tau}}(t)$ requires estimating both marginal survival function $\hat{S}(t)$ and density of survival time $\hat{f}(t)$. While Kaplan-Meier curve is commonly used to estimate $S(t)$, it is unrealistic to estimate $f(t)$ by taking the derivative of $\hat{S}(t)$, since $\hat{S}(t)$ would be a step function. Therefore, it has also been proposed to use a smoothed version of Kaplan-Meier curve. Many smoothing methods has been suggested for either counting process \citep{Hansen1983} or hazard function \citep{Wang2014}. In this paper, we smooth the estimated Kaplan-Meier curve using a Constrained Additive Model \citep{pya2021}: 

\[
\hat{S}(t) = \tilde{S}(t) + \epsilon(t)
    = \sum_{k=1}^K \zeta_k M_k(t) + \epsilon(t)
\]

where $\hat{S}(t)$ is the Kaplan-Meier estimators of marginal survival function, and smoothed survival function $\tilde{S}(t)$ is modelled as a linear combination of P-spline basis functions $M_1(t)...M_K(t)$ that are subject to the monotonicity constraint:
$\tilde{S}(t_1) > \tilde{S}(t_2)$ for $t_1 < t_2$. 

We hereafter refer to the estimator of concordance estimated using unsmoothed non-parametric  $\hat{\text{AUC}}^{I/D}(t)$ as non-parametric concordance $\hat{C}_{NP}$. The estimator by integrating $\tilde{\text{AUC}}^{I/D}(t)$, the smoothed non-parametric estimator of Incident/Dynamic AUC, will be referred to as smoothed non-parametric concordance $\hat{C}_{SNP}$. The estimator from the semi-parametric $\hat{\text{AUC}}^{I/D}(t)$, since it was introduced by \cite{hz2005}, will be referred to as the Heagerty-Zheng semi-parametric concordance $\hat{C}_{HZ}$.

In addition to estimators of Concordance based on integrating estimates of $\text{AUC}^{I/D}(t)$, we consider one additional semi-parametric estimator proposed by \cite{gh2005}
\begin{equation}
\hat{C}_{GH} = \frac{2}{n(n-1)}\sum_{i<j}\frac{I(\hat{\eta}_j-\hat{\eta}_i<0)}{1+exp(\hat{\eta}_j-\hat{\eta}_i)}+\frac{I(\hat{\eta}_i-\hat{\eta}_j<0)}{1+exp(\hat{\eta}_i-\hat{\eta}_j)}
\label{eq:c-gh}
\end{equation}

and one additional non-parametric estimator of concordance by \cite{Harrell1996}
\begin{equation}
\hat{C}_{Harrell} = 
\frac{\sum_{i<j}I(T_i<T_j)I(\hat{\eta}_i>\hat{\eta}_j)I(\delta_i=1)+I(T_i>T_j)I(\hat{\eta}_i<\hat{\eta}_j)I(\delta_j=1)}{\sum_{i<j}I(T_i<T_j)I(\delta_i=1)+I(T_i>T_j)I(\delta_j=1)} 
\label{eq:c-harrell}
\end{equation}

Similar to the semi-parametric estimator of incident sensitivity, the semi-parametric estimator of Concordance proposed by \cite{gh2005} includes terms of the form $e^{\eta}$ and excludes event status $\delta$, while Harrell's C only compares relative ranking of risk estimation to event time. We further note that Harrell's C can be inconsistent for the estimand of interest in the presence of censoring. Alternative consistent non-parametric estimators have been proposed \citep{Uno2011}, using the inverse probability weighting technique to modify the Harrell's C-index. In this paper, we focus on the original Harrell's C for simplicity of presentation and due to the fact that in our simulations and applications, the estimator proposed by \cite{Uno2011} which is consistent for truncated Concordance shows the same in- versus out-of-sample predictive performance effects with a slightly shifted distribution.

\subsection{Mechanism of Inflated Out-of-sample Estimation}
\label{subsec:mach_sim}

As described in Section~\ref{s:estimator}, the cause of the observed out-of-sample inflation of $\hat{\text{AUC}}^{I/D}(t)$ lies in the semi-parametric estimator of incident sensitivity. Remember the  estimator in Eq.\ref{eq:sens_sp} weighs observations at risk at time t by the exponential of their estimated risk scores: 

\begin{equation}
    \frac{exp(\hat{\eta}_k)}{\sum_{j}I(T_j \geq t) exp(
    \hat{\eta}_j)}
    \label{eq:sens_sp_wt}
\end{equation}
    
When an observation has a high estimated risk score $\hat{\eta}$, its corresponding weight would be very large. As a result, the semi-parametric estimator would be dominated by observations with high estimated risk, regardless of their actual event status at time $t$ or the accuracy of risk estimation. Take Figure \ref{fig-outlier} as an example. In the original dataset, the largest value of the sensitivity weight (Eq.\ref{eq:sens_sp_wt}) above is 2.8\%. Considering the sample size of 500, the contribution to the semi-parametric sensitivity estimation is similar across subjects. However, after introducing one outlier, the weight of this outlier is 99.99\%, and the second largest weight in the sample is 0.00018\%. That is, the semi-parametric estimator of incident sensitivity depends almost entirely on this outlier, when the observation itself is not even well predicted.

As a result of the inflated estimates of incident sensitivity, the Incident/Dynamic AUC will also be overestimated at the corresponding time points. 
Figure \ref{fig-outlier} provides a straightforward illustration. As the right panel shows, the semi-parametric ROC curves was driven to the upper left corner by the introduced outlier, causing the area under the curve to increase. The semi-parametric concordance, as a weighted integral of a series of overestimated AUC, would also be overestimated as a result.    

The Gonen-Heller estimator also has a semi-parametric composition, as Eq.\ref{eq:c-gh} has included the term $exp(\hat{\eta}_j-\hat{\eta}_i)$ in the denominator. Similar to the Heagerty-Zheng estimator of incident sensitivity, the Gonen-Heller estimator can be perceived as the weighted sum of all random pairs $(i,j)$. Specifically, Eq.\ref{eq:c-gh} can be rewritten as $\frac{2}{n(n-1)}\sum_{i<j}\frac{I(\hat{\eta}_j-\hat{\eta}_i<0)+I(\hat{\eta}_i-\hat{\eta}_j<0)}{1+exp(-|\hat{\eta}_i-\hat{\eta}_j|)}$. Here, the contribution to the estimation from each random pair of subject $(i,j)$ ($i<j$) is the term $\frac{1}{1+exp(-|\hat{\eta}_i-\hat{\eta}_j|)}$. It is straightforward to see that this term increases with the difference between $\hat{\eta}_i$ and $\hat{\eta}_j$. That is, pairs that have larger difference between estimated risk will drive the concordance estimate higher. However, this term has an upper bound of 1. In addition, the difference between a pair is usually less variable than the individual values. Therefore, the Gonen-Heller estimators would inflate less than the Heagerty-Zheng estimator. 

Take Figure \ref{fig-outlier} as an example. Compared to the original dataset, the introduction of one outlier with a very large risk score would lead to many pairs of observations with large difference between their risk scores. The contribution to the Gonen-Heller estimation, $\frac{1}{1+exp(-|\hat{\eta}_i-\hat{\eta}_j|)}$, will be larger for these pairs. And as a result, the concordance estimate will also be inflated. However, the degree of the inflation will be limited. In this case, this one outlier has driven the Gonen-Heller concordance estimator from 80.05\% to 80.14\%. For details about the distribution of difference of risk scores between pairs of observations, please see Figure \ref{fig:A_wt_gh} in the Appendix. 

The observations above are also  illustrated in the following sections through simulation studies and a real-world data application.

\section{Simulation Study}
\label{sec:simulation}

\subsection{Data Generating Mechanism}
\label{subsec:simulation_data_generating}

We designed a simulation study to illustrate the in- and out-of-sample behavior of semi- and non-parametric estimators introduced in Section~\ref{s:estimator} in finite samples. We simulate data under a Cox proportional hazards model framework with independent censoring.

The specific model for data generation is as follows: 
\begin{equation}
    \log \lambda(t|\boldsymbol{X}) 
    = \log \lambda_0(t) + \boldsymbol{X}^t\boldsymbol{\beta} 
    = \log(p\theta t^{p-1})+\eta, \hspace{0.3cm} t>0 
\label{eq:sim_model}
\end{equation}

$\boldsymbol{X} = [X_1, X_2, X_3]^t \in \mathbf{R}^3$ are three covariates simulated as independent $N(0,1)$ random variables, and $\boldsymbol{\beta} = (1, -1, 0.25)$ are the true values of coefficients. $\lambda_0(t) = p\theta t^{p-1}$ is the Weibull baseline hazard with $\theta=2$ and $p=2$, resulting in $E[T|\boldsymbol{X}=\boldsymbol{0}] = 0.63$. Censoring times are simulated uniformly from $(0,1)$ independent of event times, with administrative censoring for all individuals at $\tau=1$. Across simulated datasets this resulted in an average of 58.6\% of participants being censored, and an observed median event time of 0.29. Figure \ref{fig:A_time_dist} in the appendix presents the distribution of event and censoring times under this data generating mechanism. 

We simulate $1000$ datasets, each containing $N=250$ individuals to be used in model fitting  (the \textbf{training set}) and estimation of in-sample discriminative performance. Each simulated dataset contains an additional $250$ individuals simulated under the same data generating mechanism in~\eqref{eq:sim_model} whose data are not used in model fitting. These individuals represent the \textbf{testing set} and are used to evaluate out-of-sample discrimination. In- and out-of-sample discrimination estimates are compared to the true quantities whose values were estimated using methods described in the supplemental material \ref{subset:A2}. The behavior of semi- and non-parametric estimators is compared under the two scenarios mentioned in Section~\ref{s:intro}: model overfit and covariate misalignment, each of which are described in more detail below. Note that for both of these scenarios, the data generating mechanism remains as described above. The difference is in the model fit to the data (model overfit) and the distribution of $\boldsymbol{X}$ in the testing sample.

\subsubsection{Model overfit} Model overfit in our simulation study is induced by fitting a Cox model presented in Eq.~\ref{eq:sim_model} along with a set of covariates simulated independently from the outcome ($T$), censoring time ($C$), and covariates which define the true model ($\boldsymbol{X}$). That is, we simulate a new random vector $\boldsymbol{Z} \in \mathbf{R}^m$ for $m \in \{0, 20, 100\}$, with $m=0$ corresponding to no additional covariates, and $m=100$ corresponding to $100$ additional covariates. We then fit the model 
\begin{equation}
    \log \lambda(t|\boldsymbol{X}) 
    = \log \lambda_0(t) + \boldsymbol{X}^t\boldsymbol{\beta} + \boldsymbol{Z}^t\boldsymbol{\gamma} \;
\label{eq:sim_model_overfit}
\end{equation}
to the training data, using the results to obtain estimates of in- and out-of-sample discrimination. Given the sample size for the training sample and the expected number of observed events in the training sample, $m=20$ and $m=100$ represent moderate and severe overfitting issues. We note a few points here. First, the behavior we observe when fitting model~\eqref{eq:sim_model_overfit} to data generated by model~\eqref{eq:sim_model} would still occur even if the new covariates were associated with the outcome (i.e., model~\eqref{eq:sim_model_overfit} with $\boldsymbol{\gamma} \neq \boldsymbol{0}$ in the data generating mechanism). We simply chose no association ($\boldsymbol{\gamma} = \boldsymbol{0}$) as it is cleaner in that it keeps the true values of discrimination the same across all scenarios. Second, maximum likelihood estimates for $\boldsymbol{\beta}$ and $\boldsymbol{\gamma}$ (and thus risk scores) are consistent in this scenario ($[\hat{\boldsymbol{\beta}}, \hat{\boldsymbol{\gamma}}]^t \stackrel{p}{\to} [\boldsymbol{\beta}, \boldsymbol{0}]^t$).

\subsubsection{Covariate misalignment:} In the second scenario, we follow the same data generation mechanism as model~\eqref{eq:sim_model}. However in the testing sets, we introduce a small proportion ($\alpha$) of observations whose covariates are generated from a different distribution from the training set. As a result, the probability distribution function (PDF) $g$ of the covariates of one randomly-drawn subject from the testing sets could be written as the following mixed distribution:

\[
    g(\boldsymbol{X}) = (1-\alpha)g_0(\boldsymbol{X})+\alpha g_1(\boldsymbol{X})
\]

Where $g_0$ is the same distribution as covariates in the training data ($N(0, 1)$). And $g_1(\boldsymbol{X})$ represents the distribution of misaligned covariates.

Specifically, we set the proportion of misaligned observation $\alpha=0.1$. For $g_1$, we study the effect on out-of-sample behavior from two different types of change of covariate space : 1) a mean shift, where misaligned covariates are generated with a larger mean ($N(5, 1)$); and 2) a variation change, where misaligned covariates are generated with are larger variance ($N(0, 5)$).

\subsection{Results}
\label{subsec:simulation_results}

Results are presented separately by scenario (model overfit vs covariate misalignment) below. Within each scenario, we first discuss in- and out-of-sample local discrimination estimates ($\text{AUC}^{I/D}(t)$), followed by global discrimination (concordance). 

\subsubsection{Model overfit}
\label{subsec:sim_overfit}

\begin{figure}
\centering

\begin{subfigure}{\textwidth}
\centering
\includegraphics[width=\textwidth]{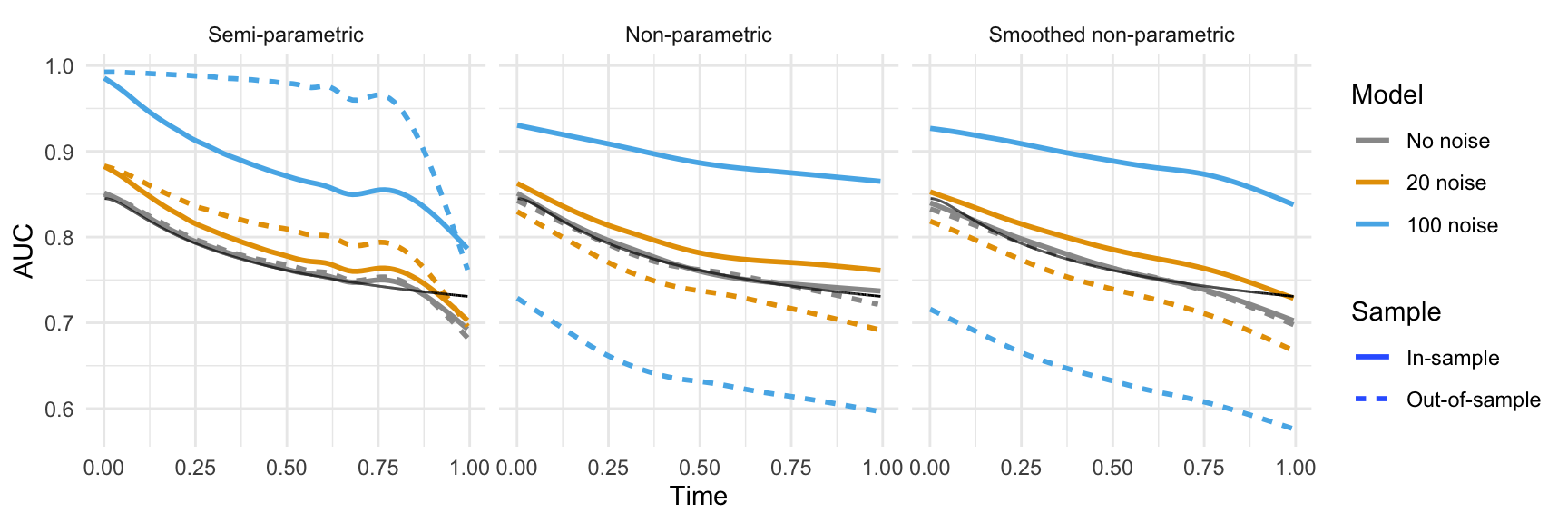}
\caption{Incident/Dynamic AUC}
\label{fig:tv_auc_overfit}   
\end{subfigure}

\begin{subfigure}{\textwidth}
\centering
\includegraphics[width=\textwidth]{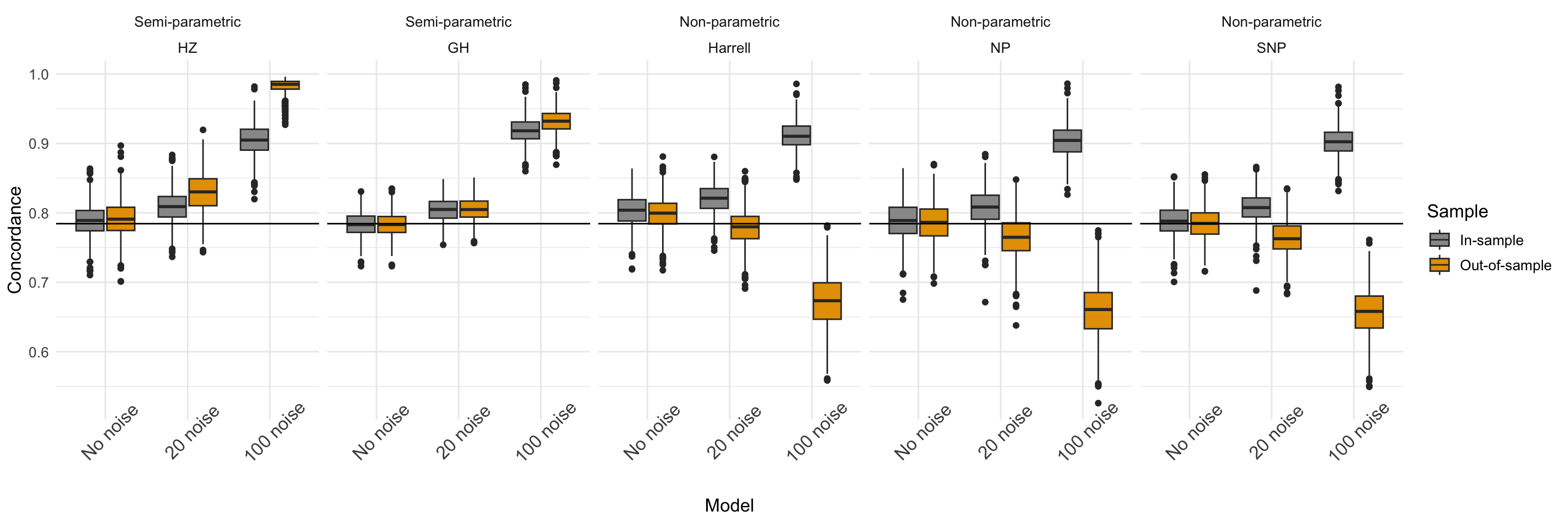}
\caption{Concordance}
\label{fig:c_overfit}   
\end{subfigure}

\caption{Behavior of estimators of model discrimination under the effect of model overfit.  Estimates of Incident/Dynamic AUC are presented in (a), smoothed across all simulations for better visualization. Solid lines represent in-sample estimates and dashed lines represent out-of-sample estimates. Line color indicates the underlying model, where grey corresponds to the correctly specified model, yellow a moderately overfit model with 20 additional signals, blue a highly overfit model with 100 additional signals. The solid black line represents true value of AUC. Estimates of concordance are presented in (b) with grey indicating in-sample and yellow out-of-sample estimates. The black horizontal line is the true value of concordance.}
\label{fig:overfit}
\end{figure}

Local discrimination, $\text{AUC}^{I/D}(t)$ is presented in Figure~\ref{fig:tv_auc_overfit}, showing trends in average discrimination, and Figure~\ref{fig:auc_box_overfit}, showing trends in the distribution of discrimination measures across simulated datasets. First consider Figure~\ref{fig:tv_auc_overfit}. Estimated discrimination under the various models fit to the data are indicated by color, with grey, yellow, and blue indicating results from models fit with no unrelated variables ($\hat{\eta}_i = \boldsymbol{X}_i^t\boldsymbol{\beta}$), 20 unrelated variables ($\hat{\eta}_i = \boldsymbol{X}_i^t\boldsymbol{\beta} + \boldsymbol{Z}_i^t\hat{\gamma}$, $\boldsymbol{Z}_i \in \mathbf{R}^{20}$), and 100 unrelated variables ($\hat{\eta}_i = \boldsymbol{X}_i^t\boldsymbol{\beta} + \boldsymbol{Z}_i^t\hat{\gamma}$, $\boldsymbol{Z}_i \in \mathbf{R}^{100}$), respectively. The black solide line represents the true values of $\text{AUC}^{I/D}(t)$. Each panel corresponds to a different estimator, with the semi-parametric estimator of \cite{hz2005}, a non-parametric estimator, and the smoothed non-parametric estimator in the left, middle, and right panels, respectively. Solid and dashed lines indicate in-sample versus out-of-sample estimated discrimination. Each line represents the average values of $\text{AUC}^{I/D}(t)$ \textit{across} simulated datasets smoothed by generalized additive model. 

In the scenario where a model is overfit to the data, we would expect the model to perform better on the training set than the testing set due to poor generalization, and the discrepancy between in-sample and out-of-sample performance should increase with the severity of model overfit. Visually, in Figure \ref{fig:tv_auc_overfit} this would correspond to solid lines being higher than the dashed lines. However, the semi-parametric estimator behaved in an opposite way. The estimates are substantially higher on the testing samples than training samples, even near perfect out-of-sample discrimination for the highly overfit model. It reveals that this estimator can overestimate the discriminative performance significantly in the presence of model overfit. However, under the correctly specified model without noise signals (gray lines), the semi-parametric estimators appear unbiased and also much smoother than the other estimators. The non-parametric estimator and smoothed non-parametric estimator, on the other hand, behaved consistently with the expectations of overfitted models, where in-sample estimates have higher values than out-of-sample estimates. When the model is not overfitted, the fully non-parametric estimator appear unbiased, while the smoothed non-parametric estimator showed slight downward bias at both ends of the follow-up period. 

Figure \ref{fig:c_overfit} visualizes the estimates of global discriminative performance across all simulations. Here, concordance estimates are summarized using boxplots, with grey boxes representing in-sample and yellow out-of-sample estimates. Each penal shows a different estimator. The two left panels, Heagerty-Zheng and Gonen-Heller estimators, are both semi-parametric. The three right panels, including Harrell's C-index, 
(unsmoothed) non-parametric and smoothed non-parametric estimators, falls into the category of non-parametric. Similar to  $\text{AUC}^{I/D}$, we expect model overfit would cause in-sample estimates to be higher than out-of-sample estimates. But the two semi-parametric estimators both have higher out-of-sample estimates than in-sample. Heagerty-Zheng estimator has a more severe out-of-sample inflation than the Gonen-Heller. On the other hand, the behavior of non-parametric estimators is more reasonable. The Harrell's c-index showed upward bias, while the others appear unbiased under the non-overfitted model.

Figure \ref{fig:auc_box_overfit} provides a more detailed comparison of the variation of different $\text{AUC}^{I/D}$ estimators, where the entire follow-up period is divided into five equal-length intervals, and the distribution of estimates within each interval is summarized by boxplots. Here, color represents the class of estimators, with grey, yellow and blue indicating the semi-parametric, non-parametric and smoothed non-parametric $\hat{\text{AUC}}^{I/D}(t)$ respectively. As it reveals, the non-parametric estimator is very unstable with the longest boxes and tails. Its value frequently fluctuates between two extremes 0 and 1. Smoothed non-parametric estimators is not as unstable, but also not as smooth as the semi-parametric estimator. All three estimators showed increasing variation over time.

\subsubsection{Covariate misalignment} 
\begin{figure}
    \centering
    \begin{subfigure}{\textwidth}
        \centering
        \includegraphics[width=\textwidth]{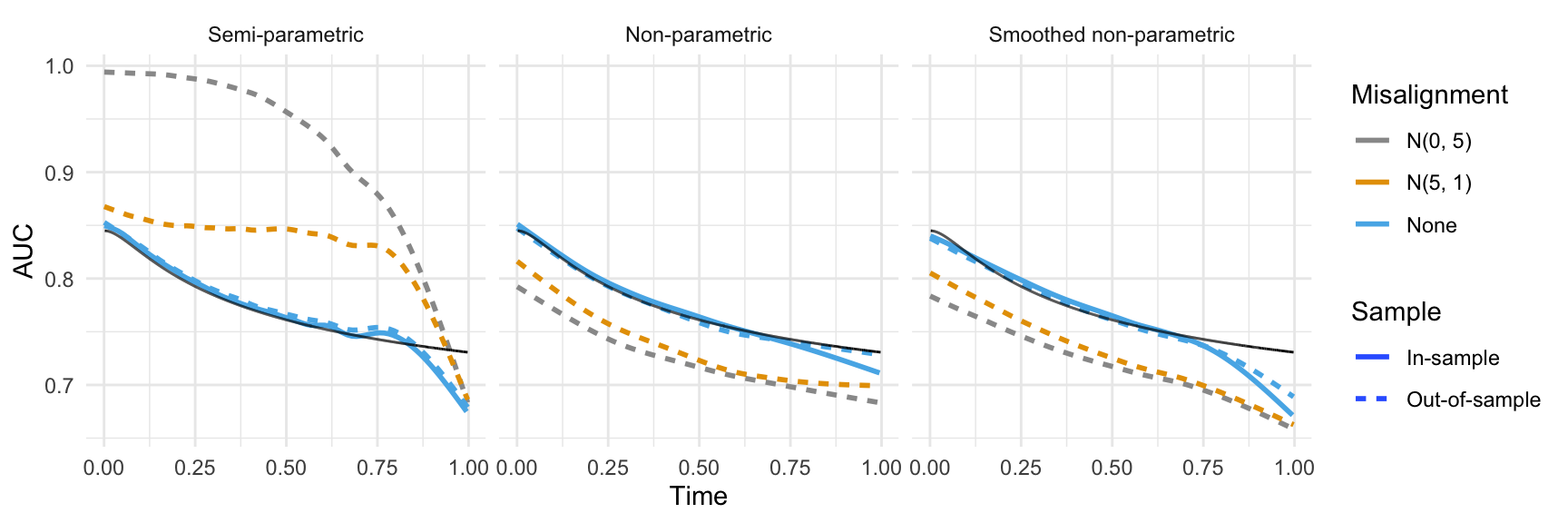}
        \caption{Incident/Dynamic AUC}
        \label{fig:tv_auc_contam}
    \end{subfigure}

    \begin{subfigure}{\textwidth}
        \centering
        \includegraphics[width=\textwidth]{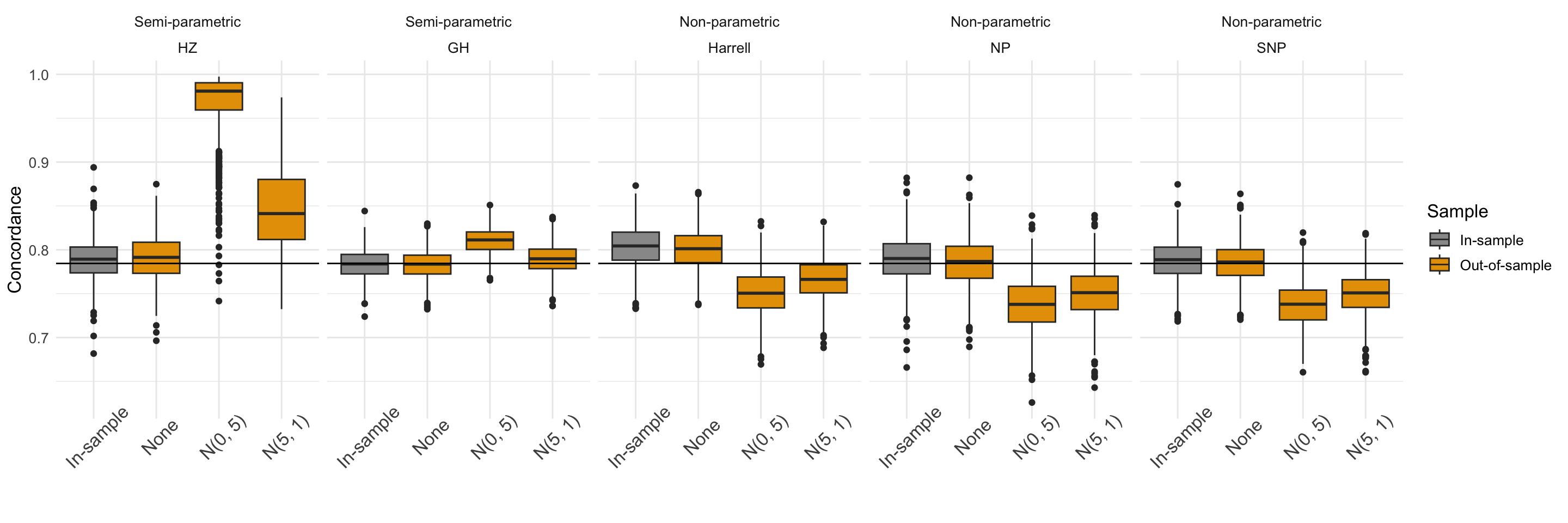}
        \caption{Concordance}
        \label{fig:c_contam}
    \end{subfigure}
    
    \caption{Behavior of estimators of model discrimination under the effect of covariate misalignment. Estimates of Incident/Dynamic AUC are presented in (a), smoothed across all simulations for better visualization. Solid lines represent in-sample estimates and dashed lines out-of-sample estimates. Line color indicates the distribution from which misaligned covariates are generated, where grey corresponds to a greater variation, yellow a mean shift, and blue the same distribution as the training sample (no misalignment). The solid black line represents true value of AUC. Estimates of concordance are presented in (b) with grey indicating in-sample and yellow out-of-sample estimates. The black horizontal line is the true value of concordance.}
    \label{fig:contam}
\end{figure}

Local and global discrimination for the covariate misalignment scenario are presented in Figures \ref{fig:tv_auc_contam} and \ref{fig:c_contam}, respectively. These figures present results in the similar form as the model overfit scenario (Section \ref{subsec:sim_overfit}). Looking at Figure \ref{fig:contam}, the same, correctly specified model with three covariates is fitted on training samples and then tested on different misaligned testing samples. Color here represents the type of covariate misalignment, with grey, yellow and blue corresponding to variation change, mean shift and no misalignment respectively. 

With the misaligned covariate space between the training and testing sets, we expect out-of-sample $\text{AUC}^{I/D}$ estimates to be lower than the in-sample ones. While the behavior of non-parametric estimators is consistent with such expectations, semi-parametric estimators show the opposite, anti-intuitive trend. In the left panel of \ref{fig:tv_auc_contam} the out-of-sample semi-parametric estimates on misaligned test samples are clearly higher than both their corresponding in-sample estimates and the true values. This inflation is more prominent when covariates have larger spread compared to mean shift. The non-parametric estimators did not suffer from over-optimistic estimation on the misaligned testing sets. On the testing sets without misalignment, the non-parametric $\hat{\text{AUC}}^{I/D}$ showed slight bias downwards at the end of the follow up period.

The concordance estimators in Figure \ref{fig:c_contam} also showed similar behavior. The semi-parametric estimators, including Heagerty-Zheng and Gonen-Heller showed higher out-of-sample than in-sample estimates on misaligned testing sets (yellow boxes). When the covariates are misaligned with a larger variation, Heagerty-Zheng estimates can be inflated close to 1, which can be interpreted as perfect discrimination. Gonen-Heller seems more robust against covariate misalignment especially when the source of misalignment is a mean shift. However, without covariate misalignment, estimators appear unbiased and smooth, except for Harrell's C with a small upward bias. 

\begin{figure}
    \centering
    \begin{subfigure}[b]{\textwidth}
        \centering
        \includegraphics[width=\textwidth]{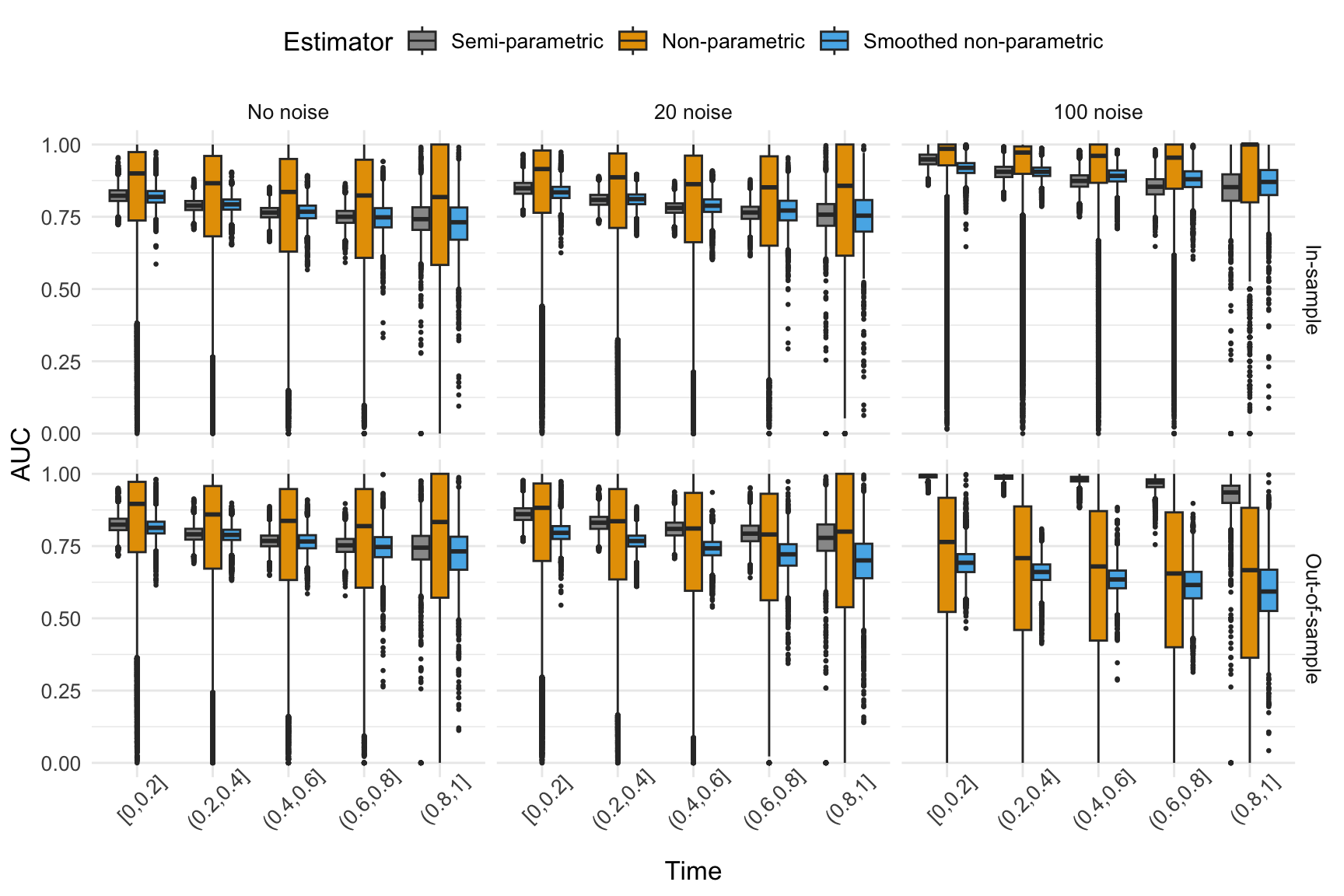}
        \caption{Model overfit}
        \label{fig:auc_box_overfit}
    \end{subfigure}
    
    \begin{subfigure}[b]{\textwidth}
        \centering
        \includegraphics[width=\textwidth]{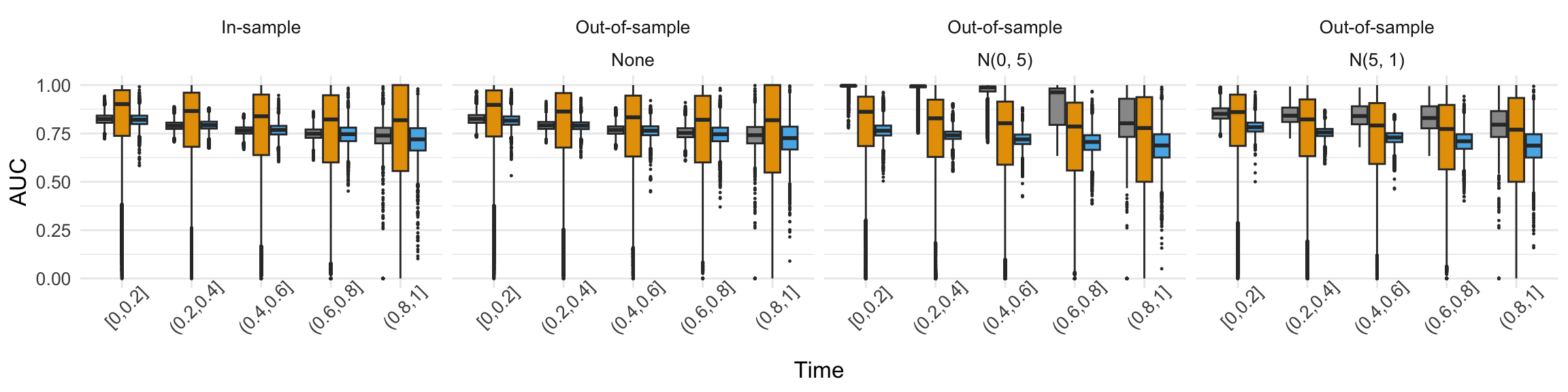}
        \caption{covariate misalignment}
        \label{fig:auc_box_contam}
    \end{subfigure}

    \caption{Comparing variability of in-sample and out-of-sample Incident/Dynamic AUC estimates. The top and bottom panels respectively reflect the effect of model overfit or covariate misalignment. The entire follow-up period is divided into five equal-length intervals, and each box represents AUC estimates in the corresponding time interval. Color of boxes represents class of estimator, with gray for semi-parametric, yellow for non-parametric and blue for smoothed non-parametric estimator.}
    \label{fig:auc_box}
\end{figure}

Figure \ref{fig:auc_box_contam} shows a comparison of the variability of $\hat{\text{AUC}}^{I/D}$. It is also clear that non-parametric estimator is much more variable than the semi-parametric one, while smoothed non-parametric has a moderate variability in between the two.


\section{Data Application}
\label{s:nhanes_data}

We illustrate the potential impact of the choice of estimator on model selection using a mortality prediction model. Recent work has used cross-validated Concordance as a criterion for evaluating the relative predictive power of data derived from wearable accelerometers versus demographic, lifestyle, and health variables known to be associated with mortality \citep{Smirnova2020, Leroux2021} and to evaluate the added value gained by considering diurnal activity patterns \citep{Cui2021} using techniques from functional regression \citep{FDAwithR, FDA}. These complex models are highly parameterized and in which overfitting is a concern. In these contexts over-inflated cross-validated estimates of Concordance could incorrectly lead to choosing a more complex model and result in substantially reduced accuracy of mortality predictions.

Using this framework as a motivation, here we use data from the National Health and Nutrition Examination Survey (NHANES) 2011-2014 to predict all-cause mortality using physical activity features derived from wearable accelerometers \citep{Leroux2019} and other participant characteristics associated with mortality as predictors. Note that the NHANES study is a multi-stage probabilistic sample from the non-institutionalized US population. Results are thus generalizable with appropriate uncertainty estimates when survey sampling methods are accounted for in regression modeling (i.e. survey weights, cluster sampling. etc). We do not account for survey design features in our application for simplicity of presentation. Specifically, eliminating the question of the impact of the distribution of survey weights on our results.

\subsection{NHANES 2011-2014 Data}
 
The analytic sample includes 3556 participants aged 50-80, with at least three days of accelerometry data with 95\% estimated wear time and complete data in covariates of interest. The total number of observed all-cause mortality events was 424, with a total of 23587.17 person-years of follow-up. Accelerometry data was processed in this spirit of the pipeline used by \cite{Leroux2021, Leroux2019} for the NHANES 2003-2006 and the UK Biobank data, respectively. The predictor vector $\boldsymbol{X_i}$ included five variables: age, Body Mass Index (BMI), Active-to-Sedentary Transition Probability (ASTP), Relative Amplitude (RA), and Total Monitor-Independent Movement Summary units (TMIMS). The latter three variables (ASTP, RA, and TMIMS) are derived from participants' wearable accelerometry data. 

\subsection{Models}
\label{subsec:appl_model}

We apply models to the data in the spirit of the ``model overfit" framework discussed in our simulation study in Section~\ref{subsec:sim_overfit}. Specifically, we consider a highly parameterized model that allows for non-linear associations and interactions between the predictor vector and risk of mortality as compared to a linear model with no interactions. The first model, an additive Cox model, is parameterized as follows: 
\begin{equation}
\log\lambda(t|\mathbf{X}_i) = \log \lambda_0(t) + f(\mathbf{X}_i) \;.
    \label{eq:acm}   
\end{equation}
The risk score in this model, $f(\mathbf{X}_i): \mathbf{R}^5 \to \mathbf{R}^1$, is a smooth function of the predictors (in this case a 5-dimensional vector) estimated using rank $200$ unpenalized thin plate regression splines \citep{wood2003} via the {\it mgcv} package \citep{wood2017} in {\it R} \citep{Rsoftware}. This model, which contains a five-dimensional smooth function, estimates 200 coefficients \textit{without} regularization. With a ratio of number of parameters to number of events of 2.12, this will tend to result in serious overfitting to the data \citep{Harrell1996}. We will refer to Model~(\ref{eq:acm}) as the "additive Cox model (ACM)". Note that the use of unpenalized regression splines is not the standard recommended in applied work. Regularization in the form of a penalty on the curvature of the estimated surface is used to control the tendency of this class of models to overfit the observed data. Here, we specifically do not use regularization/penalization to illustrate the behavior of various estimators in a clear context of model overfitting.

The second model has a simpler linear form for the risk score as follows: 
\begin{align}
    \log\lambda(t|\boldsymbol{X}_i) &= \log \lambda_0(t) + \boldsymbol{X}_i^t\boldsymbol{\beta} 
    \label{eq:lcm}
\end{align}

Given the size of the data and the number of observed events, a linear model of this form will be substantially less prone to overfitting, particularly in comparison to the ACM in~\eqref{eq:acm}. We will refer to Model~\eqref{eq:lcm} as the ``linear Cox model (LCM)".

The discriminative performance of the ACM and LCM models are evaluated using 10-fold cross-validation, using both semi- and non-parametric estimators. The former includes Heagerty-Zheng estimators of Incident/Dynamic AUC and concordance (Eq.\ref{eq:sens_sp}), as well as Gonen-Heller (Eq.\ref{eq:c-gh}) concordance. The latter includes non-parametric and smoothed non-parametric estimators of Incident/Dynamic AUC, their corresponding concordance (Eq.\ref{eq:sens_np}) (weighted by smoothed survival function) and Harrell's C-index (Eq.\ref{eq:c-harrell}). 

\subsection{Results}
\label{subsec:application_results}

\begin{figure}
    \centering
    \begin{subfigure}[b]{\textwidth}
        \centering
        \includegraphics[width=\textwidth]{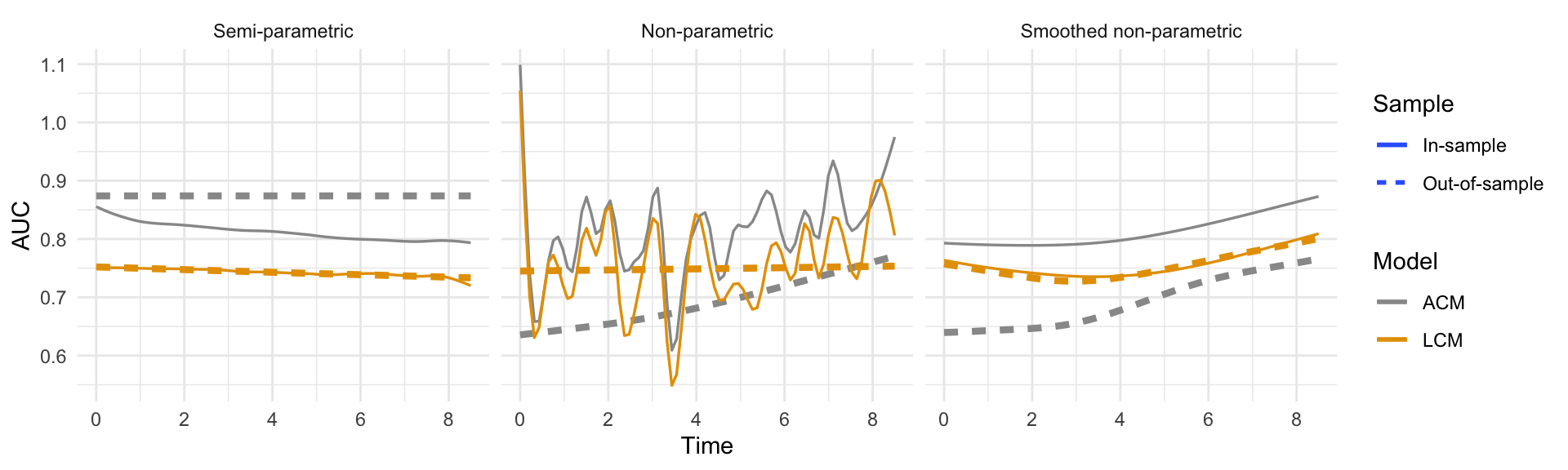}
        \caption{Incident/Dynamic AUC}
        \label{fig:appl_auc}
    \end{subfigure}
    
    \begin{subfigure}[b]{\textwidth}
        \centering
        \includegraphics[width=\textwidth]{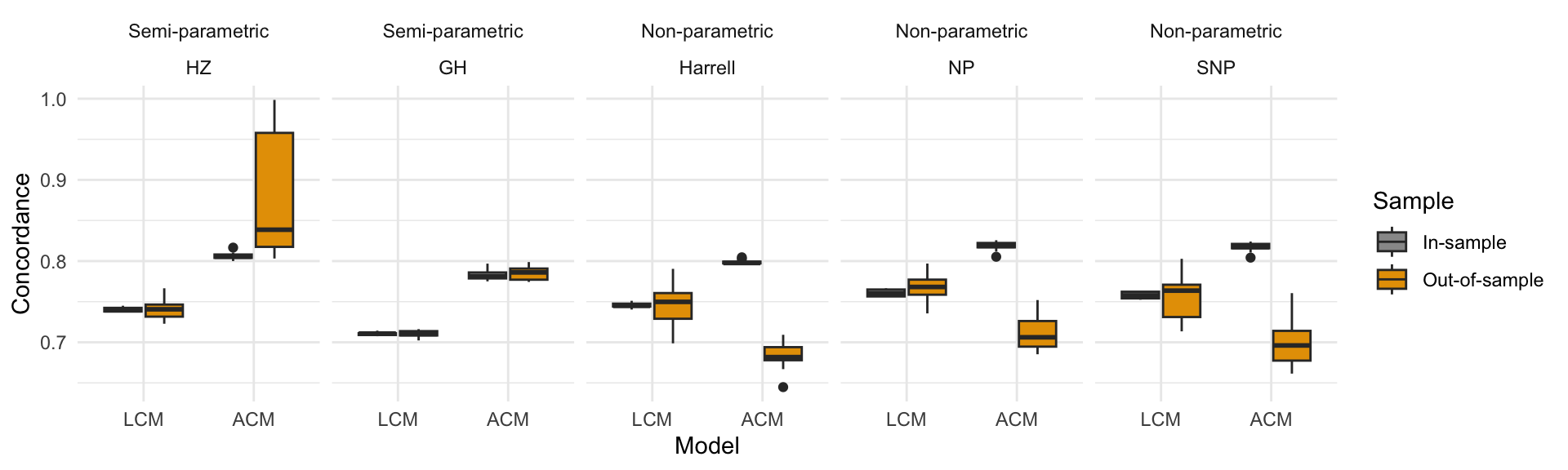}
        \caption{Concordance}
        \label{fig:appl_c}
    \end{subfigure}
    \caption{Incident/Dynamic AUC and concordance estimates on NHANES data through 10-fold cross validation. Incident/Dynamic AUC estimates are presented in (a) and smoothed over all 10 testing folds. The grey color indicates the complicated additive Cox model and yellow the simple linear Cox model; solid lines indicate in-sample and dashed lines indicate out-of-sample estimates. Concordance estimates are presented in (b), where grey indicates in-sample while yellow out-of-sample estimates.}
    \label{fig:appl}
\end{figure}

Figure \ref{fig:appl} compares the in- and out-of-sample behavior of discrimination estimators from the simpler LCM and the more complex ACM models. Results are presented in the same format as Figure~\ref{fig:tv_auc_overfit} and Figure~\ref{fig:c_overfit} for $\hat{\text{AUC}}^{I/D}(t)$ and Concordance, respectively. 

First, consider local discrimination presented in Figure~\ref{fig:appl_auc}, where grey and yellow lines indicate the additive and linear model, respectively, and solid/dashed lines indicate in- versus out-of-sample estimates, respectively. We note a few key findings. First, the in-sample behavior of the three estimators is similar, with estimates from ACM being higher than estimates from LCM (compare solid lines across panels). That the ACM has higher in-sample discrimination is to be expected due to overfitting to the training folds. However, the out-of-sample semi-parametric estimates of $\hat{\text{AUC}}^{I/D}(t)$ from the ACM are not only much higher than the estimates obtained from the LCM, but also from its own in-sample counterpart, with a striking increased out-of-sample difference in $\hat{\text{AUC}}^{I/D}(t)$ as high as 0.15. If an analyst were to make a model selection decision based on cross-validated discrimination using the semi-parametric estimator, they would incorrectly choose the more complicated ACM, which fits poorly to the data, over the simpler but better fitting LCM. In contrast, the non-parametric and smoothed non-parametric estimators showed the opposite trend when comparing in-sample (training folds) versus out-of-sample discrimination (testing folds) when evaluating the ACM. That is, the dashed grey lines presenting below the solid grey lines in the right two panels of Figure~\ref{fig:appl_auc} are consistent with the expectation that the ACM is overfitting to the data. The more parsimonious LCM shows essentially equivalent in- and out-of-sample discrimination estimates across estimators considered here. We note that the fully non-parametric estimator of in-sample discrimination (Figure~\ref{fig:appl_auc}, middle panel) appears highly non-linear for both the LCM and ACM due to its high variability and the consistency of estimates across training splits. This point is expounded upon in the Appendix. 

Figure~\ref{fig:appl_c} summarizes the in-sample (grey boxplots) and cross-validated/out-of-sample (yellow boxplots) estimates of global discriminative performance separately by estimator (panels). Note that from the perspective of model selection, for a given estimator, an analyst would choose the ACM over the LCM if the center of the yellow boxplot (average across testing splits) is higher for the ACM than the LCM. From this perspective, it is clear that both semi-parametric estimators would result in the analyst incorrectly choosing the ACM over the LCM. In contrast, using any of the non-parametric estimators would result in choosing the LCM over the ACM. Analyzing the findings in a bit more detail, we see that comparing the grey to the yellow boxplots for the LCM and ACM for a fixed estimator, the same out-of-sample inflation of estimated local discrimination ($\hat{\text{AUC}}^{I/D}(t)$) in semi-parametric estimators is also observed in Concordance. Because the Heagerty-Zheng estimator of Concordance is just a weighted integral of local discrimination, this finding for this particular estimator is unsurprising. The Gonen-Heller estimator showed higher out-of-sample estimates from ACM than LCM, though the discrepancy is much smaller than was observed in the Haegerty-Zheng estimator. The difference between mean estimates over all ten testing folds between ACM and LCM is 0.136 for Heagerty-Zheng, and 0.075 for Gonen-Heller.

\section{Discussion}
\label{sec:discussion}

\begin{table}
\centering
\begin{tabular}{llll}
\toprule
Estimator & Bias & Variability & Out-of-sample behavior\\
\midrule
\addlinespace[0.3em]
\multicolumn{4}{l}{\textbf{Incident/Dynamic AUC}}\\
\hspace{1em}Semi-parametric & Unbiased & Low & Over-optimistic\\
\hspace{1em}Non-parametric & Unbiased & High & \vphantom{1}Appropriate\\
\hspace{1em}Smoothed non-parametric & Slightly biased & Low & Appropriate\\
\addlinespace[0.3em]
\multicolumn{4}{l}{\textbf{Concordance (semi-parametric)}}\\
\hspace{1em}Heagerty-Zheng & Unbiased & Low & Over-optimistic\\
\hspace{1em}Gonen-Heller & Unbiased & Low & Over-optimistic\\
\addlinespace[0.3em]
\multicolumn{4}{l}{\textbf{Concordance (non-parametric)}}\\
\hspace{1em}Harrell & Biased upwards & Low & Appropriate\\
\hspace{1em}Non-parametric & Unbiased & High & Appropriate\\
\hspace{1em}Smoothed non-parametric & Unbiased & Low & Appropriate\\
\bottomrule
\end{tabular}

\caption{Summary of the behavior of estimators for the discriminative performance of time-to-event models}
\label{tab:sum}
\end{table}

In this paper, we compared the behavior of several semi-parametric and non-parametric estimators of both Incident/Dynamic AUC and Concordance through a simulation study in Section \ref{sec:simulation} and a case study in Section \ref{s:nhanes_data}. In the context of out-of-sample evaluation, the class of semi-parametric estimators, including the Heagerty-Zheng estimator of Incident/Dynamic AUC and concordance and the Gonen-Heller concordance, show the tendency to overestimate out-of-sample estimates of model performance, especially when the model is overfitted, or the test sample used is misaligned with a different covariate space from the training sample. We have also identified the source of this phenomena in Sections \ref{subsec:define_auc} and \ref{subsec:mach_sim} by pointing out the lack of dependence on the accuracy of risk estimation and actual event status of the incident sensitivity estimator. These estimators can thus lead analysts to incorrectly believe that a model performs much better than it actually does, resulting in their choice of an overfitted, complex model over a more appropriate, simple model. Therefore, we caution their use for model assessment, comparison and selection purposes, particularly when out-of-sample evaluation is used. On the other hand, when the model is correctly specified and sample is not misaligned, these estimators have desirable properties, such as 
consistency and smoothness. 

While the fully non-parametric estimators do not suffer from over-optimistic out-of-sample estimation, they are highly unstable due to the relatively small number of events at any given time point, motivating the need for a smoothed estimator. As such, we proposed a method for smoothing these non-parametric estimators using penalized regression splines, though other smoothers (e.g. kernel smoothing) may be used. In our simulation study, this smoothing approach works well for reducing variability, but could result in slight bias for time-dependent Incident/Dynamic AUC. We believe the bias is likely a result of heteroskedasticity of the residual process and correlation of estimates for $\text{AUC}^{I/D}(t)$ across time, which are not accounted for in classical additive models. Further methodological work for identifying a better smoother of non-parametric estimators and establishing accurate inferential procedures is needed. Although defining an optimal smoothing approach is beyond the scope of this paper, the GAM smoother used here provides a good illustration of expected in- and out-of-sample behavior for an estimator of discrimination. 

In summary, this work represents an important step forward in identifying the conditions under which various estimators of discrimination in time-to-event models are appropriate. Critically, we identified a previously ignored intrinsic flaw in a class of popular evaluation criterion for the discriminative performance of time-to-event models. Evidence are provided through simulation and case study to illustrate how such a flaw can mislead the process of model assessment and selection, and how alternative non-parametric estimators are better options for such purposes. Finally, we propose one smoothing method to mitigate the high stability of non-parametric estimators, at the cost of introducing slight bias. The behaviors and properties of all estimators are summarized in Table \ref{tab:sum} above.

\noindent {\bf{Conflict of Interest}}

\noindent {\it{The authors have declared no conflict of interest. (or please state any conflicts of interest)}}

\newpage
\appendix
\renewcommand\thefigure{A.\arabic{figure}}  
\setcounter{figure}{0}

\renewcommand{\thesection}{Appendix \Roman{section}} 

\section{Additional figures}
\label{subset:A1}

\begin{figure}
    \centering
        \includegraphics[width=0.9\textwidth]{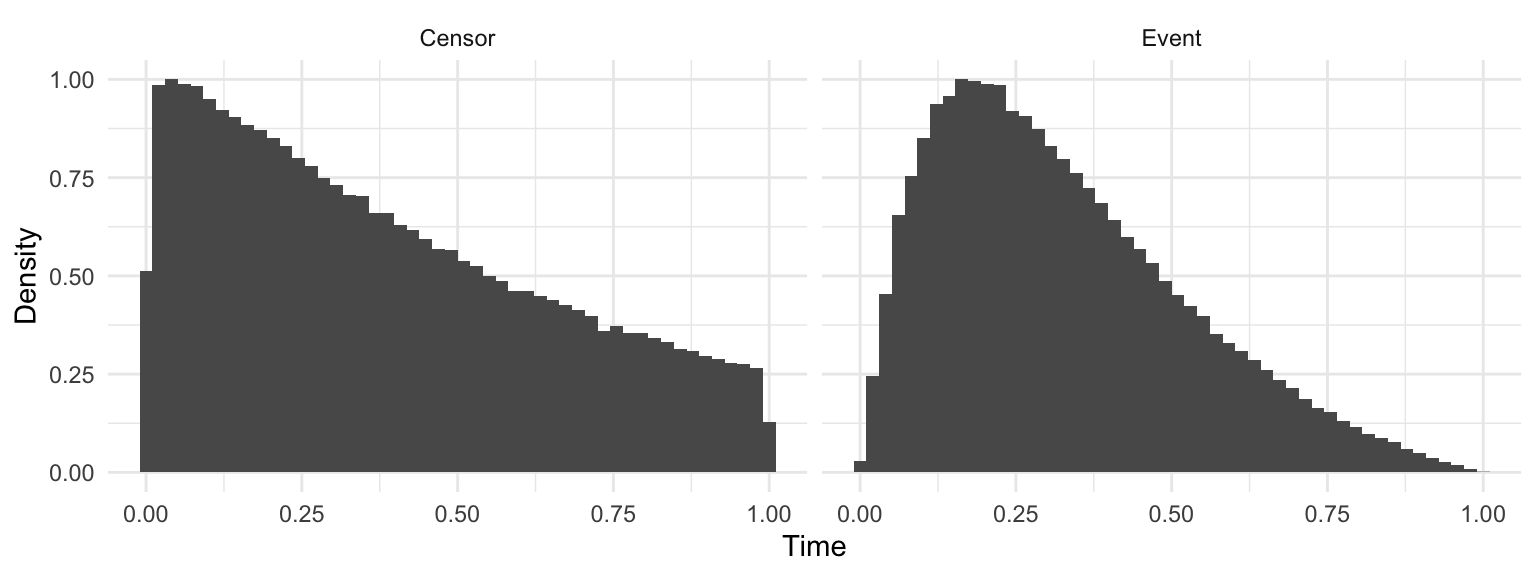}
        \caption{Distribution of event and censoring time across all simulated training sets }
        \label{fig:A_time_dist}
\end{figure}

Figure \ref{fig:A_time_dist} shows the distribution of event and censoring time across all simulated training sets in the simulation study (Section \ref{sec:simulation}).

\begin{figure}
    \centering
    \includegraphics[width=\textwidth]{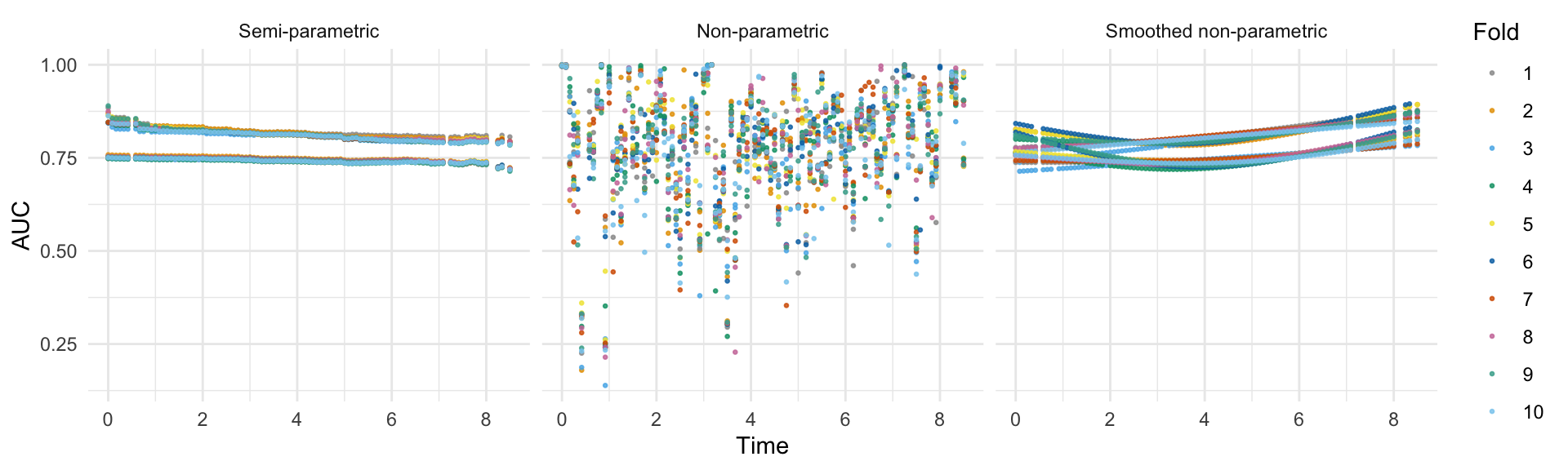}
    \caption{The unsmoothed Incident/Dynamic AUC estimates over time from the case study of NHANES data in Section 5. Color indicates the index of test fold in each iteration through the Cross Validation process.}
    \label{fig:auc_t_cv}
\end{figure}

Figure \ref{fig:auc_t_cv} shows the unsmoothed in-sample estimates of $\hat{\text{AUC}}^{I/D}(t)$ in the data application (Section \ref{s:nhanes_data}), colored by the testing fold used in each iteration during the cross validation procedure. As the middle panel shows, the in-sample estimates of non-parametric $\hat{\text{AUC}}^{I/D}(t)$ is highly numerically unstable, often fluctuating between extreme values. As a result, the smoothed value in Figure \ref{fig:appl_c} in Section \ref{subsec:application_results} is also very wiggly, especially compared to the other estimators. It is an artifact of both the high density and variability of estimates over time, and the smoothing method used for visualization.

\begin{figure}
    \centering
    \includegraphics[width=0.6\textwidth]{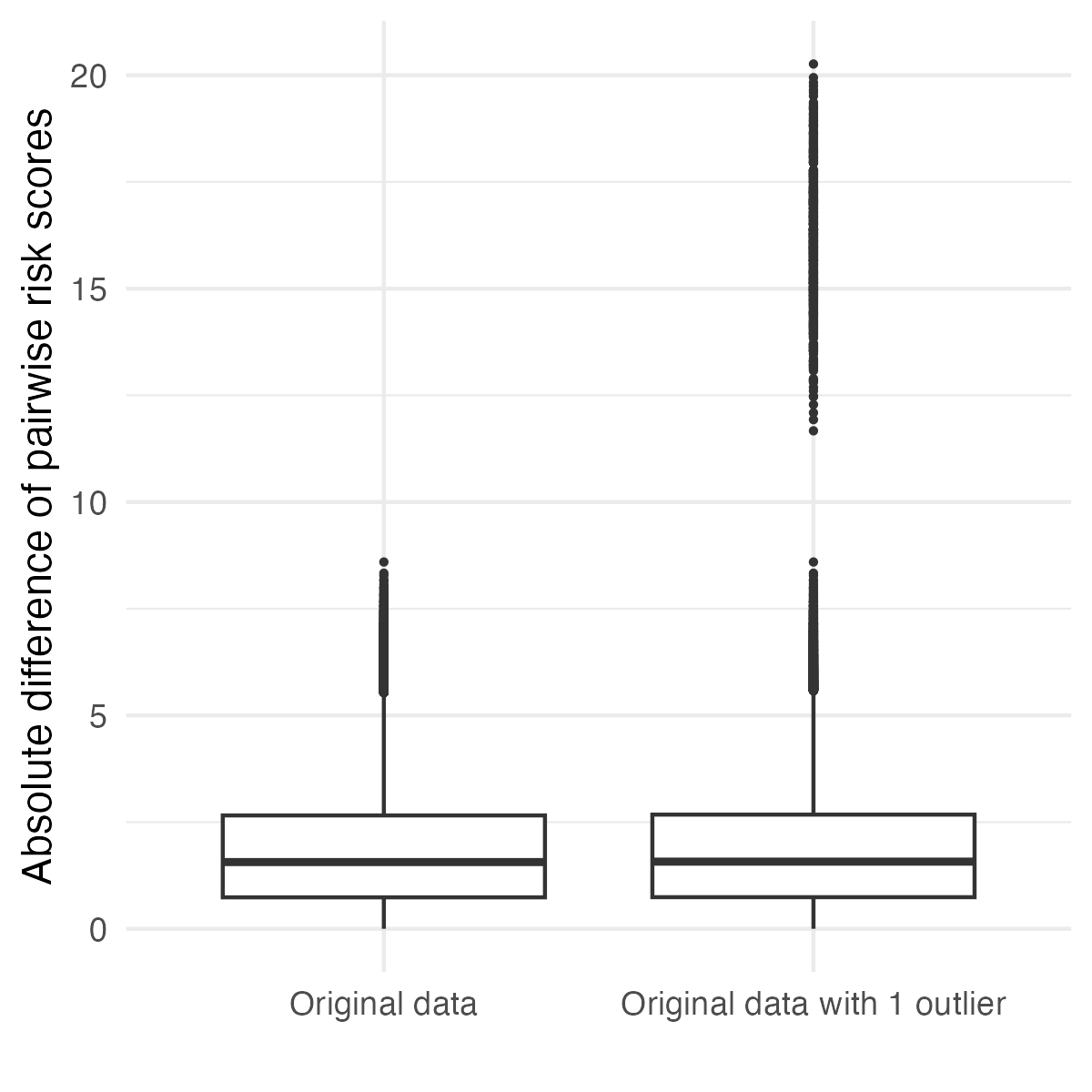}
    \caption{The change of distribution of pairwise difference between estimated risk scores after introducing one outlier in Figure \ref{fig-outlier}}
    \label{fig:A_wt_gh}
    
\end{figure}

Figure \ref{fig:A_wt_gh} is a summary of distribution of pairwise difference of estimated risk scores, calculated on the same data used to produce Figure \ref{fig-outlier}. Comparing the left and right box, the introduction of one outlying observations has introduced many pairs with large difference between risk score, causing a longer upper tail of the right box, which has driven the value of Gonen-Heller estimator higher (Section \ref{subsec:mach_sim}).

\newpage
\section{Evaluating True Incident/Dynamic AUC}
\label{subset:A2}

We can obtain true incident sensitivity and dynamic specificity under our data generating mechanism by monte-carlo integration. Specifically, consider incident sensitivity
\begin{align*}
    \text{Pr}(\eta > c|T = t) 
    &= E[1(\eta > c) | T = t] \\
    &= \int 1(\eta > c) f(\eta | t) d\eta \\
    &= \int 1(\eta > c) \frac{f(t|\eta)f(\eta)}{\int f(t|\eta) f(\eta)d\eta}d\eta
\end{align*}
Since $\eta = \bm{x}^t \bm{\beta}$ is a linear combination of normal random variables, 
$\eta$ is normally distributed. In addition, under the assumption of a Weibull baseline hazard, we can obtain 
\begin{align*}
    f( t| \eta) 
    &= \lambda(t|\eta)S(t|\eta) \\
    &= (\theta e^{\eta})p t^{p-1} e^{-(\theta e^{\eta}) t^p }
\end{align*}
We can then estimate incident sensitivity using numeric integration via, e.g., the {\it integrate()} function in {\it R}.

Next, consider dynamic specificity
\begin{align*}
    \text{Pr}(\eta \leq c|T > t) 
    &= \frac{\text{Pr}(\eta \leq c \cap T > t)}{\text{Pr}(T > t) } \\
    &= \frac{\int_t^\infty \int_{-\infty}^c f(t,\eta)d\eta dt }{\int_t^\infty [\int f(t|\eta)f(\eta)d\eta]dt } \\
        &= \frac{\int_t^\infty \int_{-\infty}^c f(t|\eta)f(\eta) d\eta dt }{\int_t^\infty [\int f(t|\eta)f(\eta)d\eta] dt} \\
\end{align*}
The double integrals involved can again be evaluated using numeric integration via, e.g., the {\it cubature::adaptIntegrate()} function in {\it R}.









\bibliography{refs}
\bibliographystyle{jds}
\nocite{*}

\end{document}